\documentclass[aps,prl,final,floatfix,a4paper,twocolumn]{revtex4-2}
\usepackage{graphicx}
\def\be{\begin{equation}}
\def\ee{\end{equation}}
\def\ba{\begin{array}}
\def\ea{\end{array}}
\def\bea{\begin{eqnarray}}
\def\eea{\end{eqnarray}}

\def\({\left(}
\def\){\right)}
\def\[{\left[}
\def\]{\right]}
\begin{document}
\title{Exotic Phase Space Dynamics Generated by Orthogonal Polynomial Self-interactions}
\author{Thokala Soloman Raju}
\email{tsraju@ifheindia.org}
\address{Department of Physics, Faculty of Science and Technology, ICFAI Foundation for Higher Education \\ %(Declared as Deemed-to-be University u/s 3 of the UGC Act 1956) \\
Dontanapally, Hyderabad 501203, Telangana, India.}
\author{T. Shreecharan}
\email{shreecharan@ifheindia.org}
\address{Department of Physics, Faculty of Science and Technology, ICFAI Foundation for Higher Education \\ %(Declared as Deemed-to-be University u/s 3 of the UGC Act 1956) \\
Dontanapally, Hyderabad 501203, Telangana, India.}
\begin{abstract}
The phase space dynamics generated by different orthogonal polynomial self-interactions exhibited in higher order nonlinear Schr\"{o}dinger equation (NLSE) are often less intuitive than those of
cubic and quintic nonlinearities. Even for nonlinearities as simple as a cubic in NLSE, the
dynamics for generic initial states shows surprising features.
In this Letter, for the first time, we identify the higher-order nonlinearities in terms of orthogonal polynomials in the generalized NLSE/GPE. More pertinently, we explicate different exotic phase space structures for three specific examples: (i) Hermite, (ii) Chebyshev, and (iii) Laguerre polynomial self-interactions. For the first two self-interactions, we exhibit that the alternating signs of the various higher-order nonlinearities are naturally embedded in these orthogonal polynomials that confirm to the experimental conditions. To simulate the phase-space dynamics that bring about by the Laguerre self-interactions, a source term should {\it necessarily} be included in the modified NLSE/GPE. Recent experiments suggest that this modified GPE captures the dynamics of self-bound quantum droplets, in the presence of external source. 
\end{abstract}
\maketitle

By now, it is a proven fact that the
interplay between dispersion and higher-order nonlinearities
can give rise to several important phenomena in optical
systems.  It is imperative that the studies of propagating solitons in NLS models involving higher-order non-Kerr nonlinearities
are much more important than the ones for the simplified NLS
equation.  For example, the cubic nonlinear Schr\"odinger equation is widely
used for descriptions of the propagation of picosecond pulses
in Kerr media \cite{8}. Nevertheless,  for the propagation of a sub-picosecond or femtosecond pulses, the higher-order effects
should be taken into account \cite{9} and one needs to describe the
problem by various generalizations of the NLS equation.  It is relevant
to mention that the measurement of fifth- and seventh-order
nonlinearities of several glasses has recently been achieved
using a spectrally resolved two-beam coupling technique \cite{11}. Recently there has been tremendous interest in analyzing higher-order Kerr response nonlinearities. These higher-order nonlinearities have been
theoretically argued to provide the main mechanism in filament
stabilization \cite{12,13}, instead of the plasma defocusing \cite{14}.
Furthermore, higher-order nonlinearities have recently been
introduced to describe the experimental observation of a
saturation (and subsequent sign inversion) of the nonlinear
correction to the refractive index at high optical intensities in gases \cite{15}. Additionally, it has recently been proved that
common optical media, such as air or oxygen, which can be
described by focusing Kerr and higher-order nonlinearities of
alternating signs, can support the propagation of new localized
light structures called fermionic light solitons and liquid light
states \cite{16}. More recently, chirped femtosecond solitons in nonlinear media
which exhibit not only third- and fifth-order nonlinearities
but even seventh- and ninth-order nonlinearity, that find practical applications have been seriously pursued \cite{triki}.

In literature, although exact solutions for the cubic-quintic and cubic-quintic-septimal-nonic NLSE have been reported, specifically, the phase space structure of the dynamical system involving cubic-quintic-septimal-nonic NLSE has not been explored, due to the hardship involved in finding the zeros of the higher order polynomials. However, in this Letter, we could successfully study the phase space structure of this dynamical system and others as we have identified the higher-order nonlinearities in terms of orthogonal polynomials---and the zeros of the orthogonal polynomials are guaranteed to be real \cite{percy}.

\begin{figure}
    \centering
    \includegraphics[width=3in,height=2in]{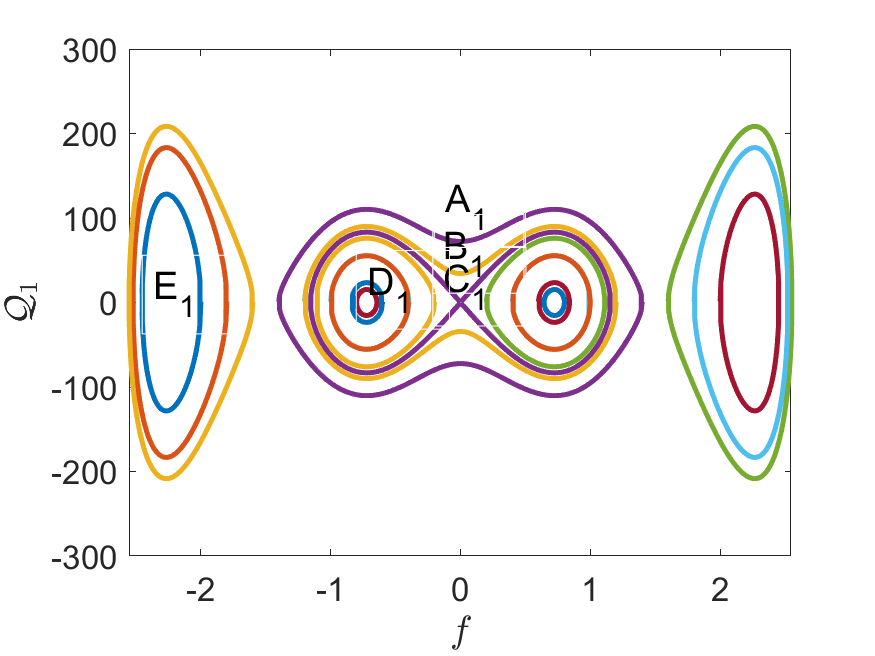}
    \caption{Phase space trajectories of Hermite self-interaction as given in Eq. (\ref{hermite1}). For a more detailed description refer to the text.}
    \label{fig:hermite}
\end{figure}

We begin our analysis by considering the NLSE with cubic-quintic-septic-nonic nonlinearities: 
\bea \nonumber
i\frac{\partial \psi}{\partial z} = \alpha_{0}\frac{\partial^2 \psi}{\partial x^2} + \alpha_{1} \psi + \alpha_{2}|\psi|^2 \psi + \alpha_{3} |\psi|^4 \psi \\ +\alpha_{4}|\psi|^6 \psi + \alpha_{5} |\psi|^8 \psi \label{eom1}
\eea
This equation is associated with $\frac{\delta \mathcal{L}}{\delta \psi^\ast}=0$, in which the Lagrangian density is given by
\bea \nonumber
\mathcal{L} = i (\psi \psi_z^\ast - \psi_z^\ast \psi) + \alpha_{0}|\psi_x|^2 +   \alpha_{1}|\psi|^2 +\alpha_{2} |\psi|^4  \\ + \alpha_{3} |\psi|^6  
+\alpha_{4} |\psi|^8 + \alpha_{5} |\psi|^{10}  \ . \label{lag1}
\eea
In order obtain the exact solitary wave solutions of Eq. (\ref{eom1}) we
consider a solution of the form
\be
\psi(x,z)=f(\xi)e^{i[\chi(\xi)-\Omega z]} \label{ansatz1}
\ee
where $\xi=x-vz$ and $\Omega$  indicates the propagation constant and $v$ represents speed of the wave. Substituting this ansatz into Eq. (\ref{eom1})  yields the following set of coupled equations:
\bea  \nonumber
 \Omega f -\chi^{\prime}f + \alpha_{0} f^{\prime\prime}-\alpha_{0}f{\chi^{\prime}}^ {2}+ \alpha_{1}f+\alpha_{2}f^{3} + \\ \label{real1}\alpha_{3} f^5  +\alpha_{4}f^7  + \alpha_{5} f^9 =0,\\
 -vf^{\prime}+2\alpha_{0}f^{\prime}\chi^{\prime}+\alpha_{0}f\chi^{\prime\prime}=0 \label{imaginary1}.
\eea  
Solving Eq. (\ref{imaginary1}), we get 
\be 
\chi^{\prime}=\frac{v}{2\alpha_{0}}-\frac{C}{2f^{2}},
\ee 
where $C$ is an integration constant. In order that the phase is independent of $\psi$, we put $C=0.$ Then Eq. (\ref{real1}) takes the form, after invoking the constant chirp into it:
\bea \nonumber 
\alpha_{0}f^{\prime\prime}+(\Omega + \alpha_{1} -\frac{v}{2\alpha_{0}}-\frac{v^{2}}{4\alpha_{0}}) f +\alpha_{2}f^{3} + \alpha_{3} f^5 \\ +\alpha_{4}f^7  + \alpha_{5} f^9 =0. \label{elliptic1}
\eea
We choose the coefficients of various powers of $f$ as in the physicists version of Hermite polynomials. This identification not only represents various orders of higher order nonlinearities but also the alternating signs of them that is required for their experimental observations in systems such as  response to ultrashort laser
pulses of common optical media, and filament stabilization \cite{david}. 
In view of this, with brevity Eq. (7) can be rewritten as
\be 
\frac{d^2 f}{d \xi^2} = \mathcal{P}_{9}(f).\label{poleq}
\ee
Herein, for $\mathcal{P}_{9}(f)$ we take the Hermite polynomial and Chebyshev polynomials as examples \cite{szego}.
We describe below the phase-space dynamics of these two orthogonal polynomials. We start with Hermite polynomial self-interaction.\\
\noindent {\it Hermite self-interactions}.----
For this type of self-interaction, Eq. (\ref{poleq}) takes the form
\be 
\frac{d^2 f}{d \xi^2} = H_{9}(f) \label{hermite1}
\ee
where $H_{n}(f)$ is the nth-order Hermite polynomial.
Finding the first energy integral, we obtain
\bea \nonumber
(\frac{d f}{d \xi})^{2} = 30240 f^{2} - 40320 f^4  + 16128 f^6  -2304 f^8 \\  
+ 102.4 f^{10} -3024 + 2 \mathcal{E}_1. \label{energyhermite}
\eea
Equation (10) describes
the evolution of the anharmonic oscillator with potential
\bea \nonumber
\mathcal{V}(f)= -30240 f^{2} + 40320 f^4  - 16128 f^6 \\ + 2304 f^8  - 102.4 f^{10} .
\eea
Equation (9) can be  represented by two equivalent first-order
differential equations given by
\bea 
\frac{d f}{d\xi}=\mathcal{Q}_{1}=p_{1}(f,\mathcal{Q}_{1}),\\
{\rm and }~~~~~
\frac{d \mathcal{Q}_{1}}{d\xi}=H_{9}(f)=q_{1}(f,\mathcal{Q}_{1}).
\eea 
It is rather straightforward to show that Eqs. (12) and (13) form
a Hamiltonian system, satisfying the canonical equations
\bea 
\frac{d f}{d\xi}=\frac{\partial \mathcal{H}_{1}}{\partial \mathcal{Q}_{1}};~~~\frac{d \mathcal{Q}_{1}}{d\xi}=-\frac{\partial \mathcal{H}_{1}}{\partial f},
\eea 
with the integration constant $\mathcal{E}_{1}$ given in Eq. (\ref{energyhermite}) as the
Hamiltonian $\mathcal{H}_{1}$  of the dynamical system described by Eq. (14).
We obtained nine equilibrium points $(f_{i}^{*},\mathcal{Q}_{i}^{*}) (i = 1, ..., 9)$ of Eq. (14) by solving $\frac{df}{d\xi}=0$ and $\frac{d \mathcal{Q}_{1}}{d \xi}=0.$ These are:
\bea \nonumber
(0,0) \ ,\quad (\pm 0.7236,0) \ ,\quad (\pm 1.4686,0) \ , \\
(\pm 2.2666,0) \ ,\quad (\pm 3.1910,0) \ .
\eea
Using the general criteria for the linear stability analysis, we have checked from the eigenvalues of the Jacobian matrix that the equilibrium point $(0, 0)$ is a saddle or hyperbolic type
and $(\pm 0.7236,0) \ ,\quad (\pm 1.4686,0) \ , \\
(\pm 2.2666,0) \ ,\quad (\pm 3.1910,0)$ represent elliptic equilibrium points, also
called centers. Now we display the numerically computed phase trajectories using Chebfun package \cite{chebfun},  near
the equilibrium points 
\bea \nonumber
(- 0.7236,0), (- 1.4686,0), (- 2.2666,0), (- 3.1910,0) \\ \nonumber
(0,0), (0.72356,0), ( 1.4686,0), (2.2666,0), (3.1910,0)
\eea
in Fig. (\ref{fig:hermite}). The first four points correspond to centers, the fifth point signifies a saddle, and the last four points correspond to
centers, respectively, as expected. As can be seen from this phase space diagram, between
the external (phase curves $A_1$ and $B_1$) and internal closed orbits (curves
$D_1$) there exists a phase path which joins the equilibrium points
$(0, 0)$ to itself [denoted by paths $C_1$ in Fig.(\ref{fig:hermite})] and is a form of
separatrix known as a homoclinic orbit. The separatrix is the
curve that separates the phase space into two distinct areas.
There are also well separated closed orbits [curves $E_1$] on both sides of the curves $A_1$,$B_1$, and $D_1$. 
\\

\noindent {\it Chebyshev self-interactions}.---- Here we describe the phase-space dynamics that arise due to the Chebyshev self-interactions. For this type of self-interaction, Eq. (8) takes the form
\be \label{chebyt1}
\frac{d^2 f}{d \xi^2} =- T_{9}(f)=-(256 f^9 - 576 f^7 + 432 f^5 - 120 f^3 + 9 f),
\ee
where $T_{9}(f)$ is the 9th-order Chebyshev type-I polynomial.
Finding the first energy integral, we obtain
\be
(\frac{d f}{d \xi})^{2} = -(51.2 f^9 - 144.5 f^7 + 144 f^5 - 60 f^3 + 9 f + 2\mathcal{E}_2).
\ee 
Equation (17) describes
the evolution of the anharmonic oscillator with potential
\be 
\mathcal{V}(f)= -(51.2 f^9 - 144.5 f^7 + 144 f^5 - 60 f^3 + 9 f).
\ee
Equation (16) can be  represented by two equivalent first-order
differential equations given by
\bea 
\frac{d f}{d\xi}=\mathcal{Q}_{2}=p_{2}(f,\mathcal{Q}_{2}),\\
{\rm and }~~~~~
\frac{d \mathcal{Q}_{2}}{d\xi}=T_{9}(f)=q_{2}(f,\mathcal{Q}_{2}).
\eea 
We note that Eqs. (19) and (20) form
a Hamiltonian system, satisfying the canonical equations
\bea 
\frac{d f}{d\xi}=\frac{\partial \mathcal{H}_{2}}{\partial \mathcal{Q}_{2}};~~~\frac{d \mathcal{Q}_{2}}{d\xi}=-\frac{\partial \mathcal{H}_{2}}{\partial f},
\eea 
with the integration constant $\mathcal{E}_{2}$ given in Eq. (17) as the
Hamiltonian $\mathcal{H}_{2}$  of the dynamical system described by Eq. (16).
We obtained nine equilibrium points $(f_{i}^{*},\mathcal{Q_{i}}^{*}) (i = 1, ..., 9)$ of Eq. (21) by solving $\frac{df}{d\xi}=0$ and $\frac{d \mathcal{Q}_{2}}{d \xi}=0.$ These are:
\bea \nonumber
(0,0) \ ,\quad (\pm 0.3420,0) \ , \\ 
(\pm 0.6428,0) \ , \quad (\pm 0.8660,0) \ ,\quad (\pm 0.9848,0) \ .
\eea
In much the same way, we can choose the self-interactions to be of Chebyshev type-II i.e, $U_{9}(f)$ and study the phase-space dynamics. It is very interesting to note that the dynamics that arise due to this orthogonal polynomial is quite different from the dynamics of Chebyshev type-I, as depicted in Fig. (3).
\begin{figure}
    \centering
    \includegraphics[width=3in,height=2in]{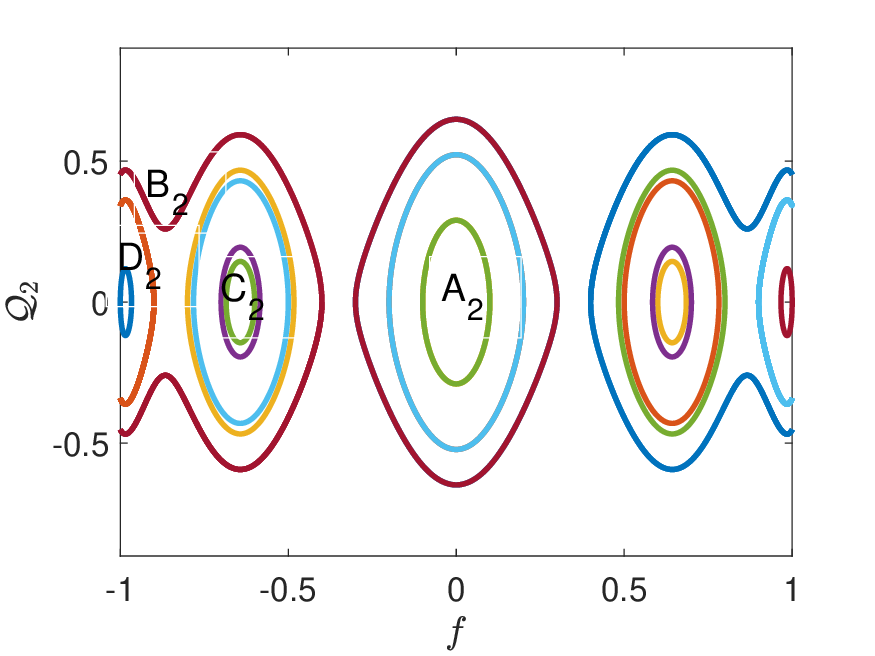}
    \caption{Phase space trajectories of Chebyshev-type I self-interaction as given in Eq. (\ref{chebyt1}). For a more detailed description refer to the text.}
    \label{fig:chebyt1}
\end{figure}

\begin{figure}
    \centering
    \includegraphics[width=3in,height=2in]{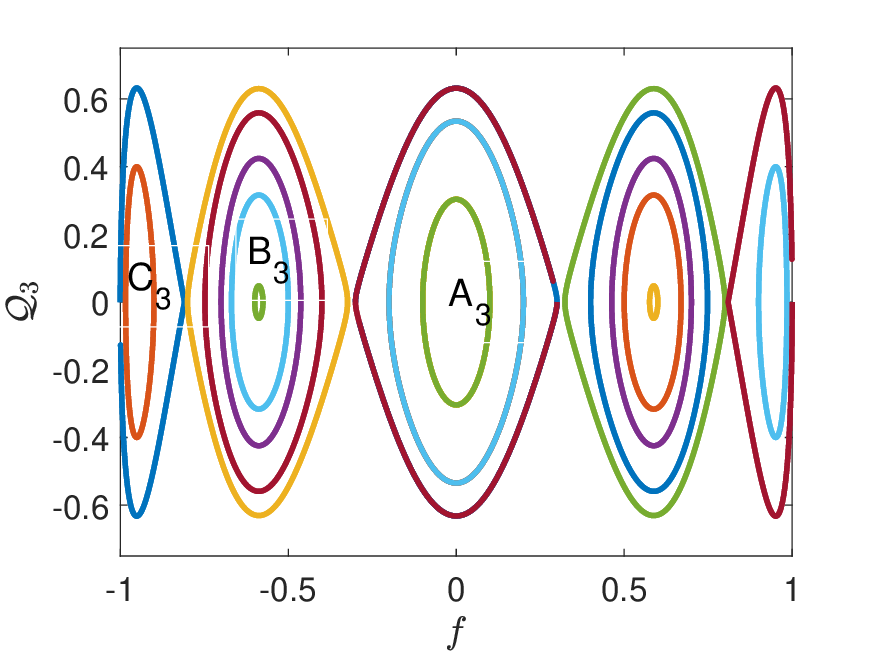}
    \caption{Phase space trajectories of Chebyshev-type II self-interaction.}
    \label{fig:chebyt2}
\end{figure}
Utilizing the linear stability analysis, we have checked from the eigenvalues of the Jacobian matrix that the equilibrium point $(0, 0)$ is a saddle or hyperbolic type
and $(\pm 0.3420,0), (\pm 0.6428,0), (\pm 0.8660,0), (\pm 0.9848,0)$ represent elliptic equilibrium points, also
called centers. In Fig.(\ref{fig:chebyt1}), we portray different phase trajectories that we have computed numerically using Chebfun package \cite{chebfun} near
the following equilibrium points 
\bea \nonumber
(- 0.3420,0), (- 0.6428,0), (- 0.8660,0), (- 0.9848,0), \\ \nonumber
(0,0), ( 0.3420,0), (0.6428,0), (0.8660,0), (0.9848,0) .
\eea The first four points correspond to centers, the fifth point signifies a saddle, and the last four points correspond to
centers, respectively, as expected. The phase curve $A_2$ represents a well separated closed orbit. $B_2$ indicates a closed external orbit that encloses the internal closed orbits $C_2$ and $D_2$ which are again well separated. There is no separatrix that joins them. The same is true for the phase orbits on the right side of the phase curves $A_2.$
Following the same methods, in Fig.(\ref{fig:chebyt2}) we display the phase space orbits corresponding to the self-interactions in the form of Chebyshev-II i.e., $U_{9}(f)$. The phase orbits $A_3$, $B_3$, and $C_3$ all are separate elliptic orbits. Here also there is no separatrix that connects these various phase orbits.\\

\begin{figure}
    \centering
    \includegraphics[width=3in,height=2in]{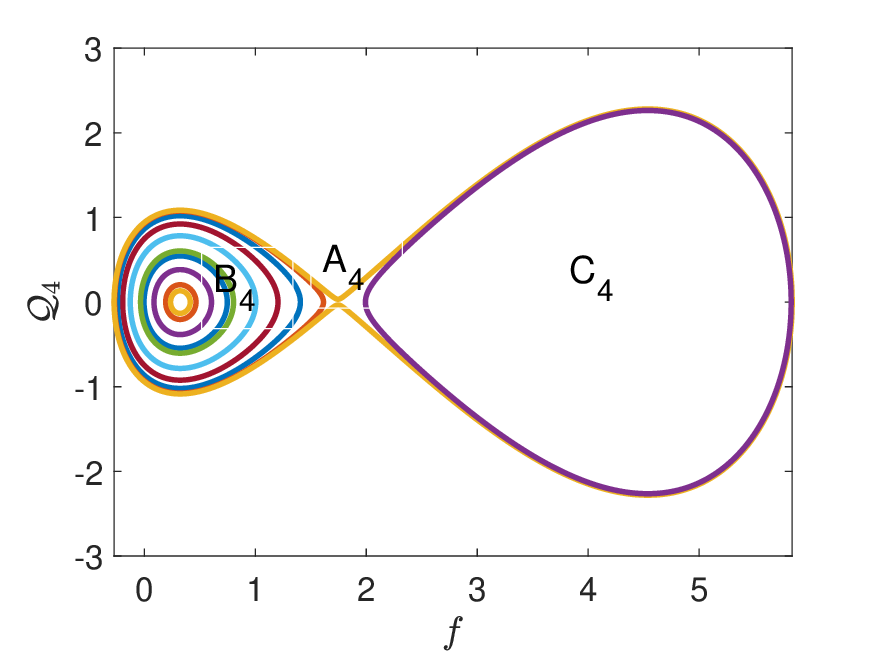}
    \caption{Phase space trajectories of Laguerre self-interaction as given in Eq. (\ref{laguerre}). For a more detailed description refer to the text.}
    \label{fig:laguerre}
\end{figure}

\noindent {\it Laguerre self-interactions}.---- As a third example, we describe the phase-space dynamics that arise due to the Laguerre self-interactions. The fact  that the Laguerre polynomial comes with a constant term in its expansion necessitates us to introduce an external source in the ensuing equation of motion. Spurred on by this fact, we start with a NLSE/GPE with an external source:
\bea \nonumber
i\frac{\partial \psi}{\partial z} = \alpha_{0}\frac{\partial^2 \psi}{\partial x^2} + \alpha_{1} \psi + \alpha_{2}|\psi| \psi \\ + \alpha_{3} |\psi|^2 \psi +\alpha_{4}|\psi|^3 \psi + \alpha_{5} e^{i\varphi(x,z)}
\eea
This equation is associated with $\frac{\delta \mathcal{L}}{\delta \psi^\ast}=0$, in which the Lagrangian density is given by
\bea \nonumber
\mathcal{L} = i (\psi \psi_z^\ast - \psi_z^\ast \psi) + \alpha_{0}|\psi_x|^2 +   \alpha_{1}|\psi|^2 +\alpha_{2} |\psi|^3 \\ + \alpha_{3} |\psi|^4 +\alpha_{4} |\psi|^5 + \alpha_{5} e^{i\varphi(x,z)}\psi^{\ast}  \ .
\eea
To obtain the exact solitary wave solutions of Eq. (23) we use the same ansatz as given in Eq. (3) that yields the following set of coupled equations:
\bea \nonumber  
 \Omega f -\chi^{\prime}f + \alpha_{0} f^{\prime\prime}-\alpha_{0}f{\chi^{\prime}}^ {2}+\alpha_{1}f+\alpha_{2}f^{2} + \alpha_{3} f^3  \\ 
 + \alpha_{4}f^4  + \alpha_{5} =0,\\
 -vf^{\prime}+2\alpha_{0}f^{\prime}\chi^{\prime}+\alpha_{0}f\chi^{\prime\prime}=0.
\eea  
Again by solving Eq. (26), we get 
\be 
\chi^{\prime}=\frac{v}{2\alpha_{0}}-\frac{C}{2f^{2}},
\ee 
where $C$ is an integration constant. In order that the phase is independent of $\psi$, we put $C=0.$ Then Eq. (25) takes the form:
\be 
\alpha_{0}f^{\prime\prime}+(\Omega+\alpha_{1} -\frac{v}{2\alpha_{0}}-\frac{v^{2}}{4\alpha_{0}}) f +\alpha_{2}f^{2} + \alpha_{3} f^3  +\alpha_{4}f^4  + \alpha_{5}  =0.
\ee 
We choose the coefficients of various powers of $f$ as in the physicists version of Laguerre polynomials. This identification not only represents various higher order nonlinearities in modified GPE but also captures the dynamics of self-bound quantum droplets \cite{cabrera}, in the presence of an external source. The external source here models the coupling of a reservoir of Bose-Einstein condensed atoms to the waveguide \cite{paul,soloman}.
For this type of self-interation, Eq. (28) takes the form
\be \label{laguerre}
\frac{d^2 f}{d \xi^2} = L_{4}(f)=\frac{1}{24}f^{4}-\frac{2}{3}f^{3}+3f^{2}-4f+1
\ee
where $L_{n}(f)$ is the nth-order Laguerre polynomial.
Finding the first energy integral, we obtain
\be
(\frac{d f}{d \xi})^{2} = \frac{1}{60} f^{5} - \frac{1}{3} f^4  + 2 f^3  -4 f^2  + 2 f +2\mathcal{E}_4.
\ee 
Equation (30) describes
the evolution of the anharmonic oscillator with the potential
\be 
\mathcal{V}(f)= - \frac{1}{60} f^{5} + \frac{1}{3} f^4  - 2 f^3  + 4 f^2  - 2 f .
\ee
Equation (29) can be  represented by two equivalent first-order
differential equations given by
\bea 
\frac{d f}{d\xi}=\mathcal{Q}_{4}=p_{4}(f,\mathcal{Q}_{4}),\\
{\rm and }~~~~~
\frac{d \mathcal{Q}_{4}}{d\xi}=L_{4}(f)=q_{4}(f,\mathcal{Q}_{4}).
\eea 
It can be shown that Eqs. (32) and (33) form
a Hamiltonian system, satisfying the canonical equations
\bea 
\frac{d f}{d\xi}=\frac{\partial \mathcal{H}_{4}}{\partial \mathcal{Q}_{4}};~~~\frac{d \mathcal{Q}_{4}}{d\xi}=-\frac{\partial \mathcal{H}_{4}}{\partial f},
\eea 
with the integration constant $\mathcal{E}_{4}$ given in Eq. (30) as the
Hamiltonian $\mathcal{H}_{4}$  of the dynamical system described by Eq. (29).
We obtained four equilibrium points $(f_{i}^{*},\mathcal{Q}_{i}^{*}) (i = 1, ..., 4)$ of Eq. (34) by solving $\frac{df}{d\xi}=0$ and $\frac{d \mathcal{Q}_{4}}{d \xi}=0.$ These are:
\be
( 0.3225,0) \quad (1.7458,0) \quad (4.5366,0) \quad (9.3951,0)
\ee
Again use was made of the general criteria for the linear stability analysis in order to check from the eigenvalues of the Jacobian matrix that the following zeros of the $L_4$:  $( 0.3225,0) \quad (1.7458,0) \quad (4.5366,0) \quad (9.3951,0) $ to represent elliptic equilibrium points, also
called centers. We have numerically integrated Eq. (29) using Chebfun package \cite{chebfun} and computed the phase trajectories near
the equilibrium points $ ( 0.3225,0) \quad (1.7458,0) \quad (4.5366,0) \quad (9.3951,0) .$ We depict these insights  
in Fig.(\ref{fig:laguerre}).  All these four points correspond to centers. The external phase curve $A_4$ represents a separatrix-like curve in disguise. The phase curves $B_4$ and $C_4$  represent a well separated closed orbits. 
\\

\noindent {\it Conclusion}.----
To conclude, we have identified the self-interactions of NLSE/GPE signifying higher order nonlinearities in terms of Hermite, Chebyshev, and Laguerre polynomials.
The fact that the zeros of these ensuing orthogonal polynomials are \textit{guaranteed} to be real really propelled us to identify these higher order nonlinearities in terms of these polynomials to unravel the rich phase space structures. Additionally these identifications resulted in alternating signs of the various higher-order nonlinearities that confirm to the experimental conditions. Driven systems like NLSE/GPE with an external source aptly model wave propagation
through asymmetric twin core fibers and self-bound quantum droplets in BEC. Indeed the identification of various higher order nonlinearities emanating from Laguerre polynomials necessitated the inclusion of an external source in the equation of motion. This external source describes the coupling of a reservoir of BEC atoms to the waveguide. We hope that the present study may stimulate to model higher order self-interactions in terms of other orthogonal and exceptional orthogonal polynomials \cite{kamran1,kamran2} that may find applications in both optics and nonlinear dynamics.


\begin{thebibliography}{99}

\bibitem{8} G. P. Agrawal, Nonlinear Fiber Optics, 5th ed. (Academic Press,
New York, 2013).

\bibitem{9} Y. Tao and J. He, Phys. Rev. E {\bf 85}, 026601 (2012).

\bibitem{11}  Y.-F. Chen, K. Beckwitt, F. W. Wise, B. G. Aitken, J. S. Sanghera,
and I. D. Aggarwal, J. Opt. Soc. Am. B {\bf 23}, 347 (2006).

\bibitem{12}  L. Berge, S. Skupin, R. Nuter, J. Kasparian, and J.-P. Wolf, ´ Rep.
Prog. Phys. {\bf 70}, 1633 (2007).
\bibitem{13}  A. Couairon and A. Mysyrowicz, Phys. Rep. 441, 47 (2001)

\bibitem{14}  P. Bejot, J. Kasparian, S. Henin, V. Loriot, T. Vieillard, E. Hertz,
O. Faucher, B. Lavorel, and J.-P. Wolf, Phys. Rev. Lett. {\bf 104},
103903 (2010).

\bibitem{15}  V. Loriot, E. Hertz, O. Faucher, and B. Lavorel, Opt. Express
{\bf 17}, 13429 (2009); {\bf 18}, 3011 (2010).

\bibitem{16}  D. Novoa, H. Michinel, and D. Tommasini, Phys. Rev. Lett. {\bf 105},
203904 (2010).

\bibitem{triki} H. Triki, K. Porsezian, A. Choudhuri, and P. T. Dinda,
Phys. Rev. A {\bf 93}, 063810 (2016).

\bibitem{percy} P. Deift, \textit{Orthogonal polynomials and random matrices: A Riemann-Hilbert approach} (American Mathematical Society, Rhode Island, 1999).

\bibitem{david} D. Novoa, H. Michinel, and D. Tommasini, Phys. Rev. Lett. \textbf{105}, 203904 (2010).

\bibitem{szego} G. Szeg\"{o}, \textit{Orthogonal polynomials} (American Mathematical Society, Rhode Island, 1975).

\bibitem{chebfun} T. A. Driscoll, N. Hale, and L. N. Trefethen, editors, \textit{Chebfun Guide} (Pafnuty Publications, Oxford, 2014).

\bibitem{cabrera} P. Cheiney, C. R. Cabrera, J. Sanz, B. Naylor, L. Tanzi, and L. Tarruell, Phys. Rev. Lett. \textbf{120}, 135301 (2018).

\bibitem{paul} T. Paul, K. Richter, and P. Schlagheck, Phys. Rev. Lett. \textbf{94}, 020404 (2005).

\bibitem{soloman} T. Soloman Raju, T. Shreecharan, and S. S. Ranjani, IEEE Journal of Quantum Electronics, \textbf{58}, 6100405 (2022).

\bibitem{kamran1} D. G\'{o}mez-Ullate, N. Kamran, and R. Milson, J. Math. Anal. Appl. \textbf{359}. 352 (2009).

\bibitem{kamran2} D. G\'{o}mez-Ullate, N. Kamran, and R. Milson, J. Approx. Theory \textbf{162}, 987 (2010).

\end{thebibliography}
\end{document}